\documentclass[aps,pra,superscriptaddress,twocolumn]{revtex4}
\usepackage{graphicx,amssymb,amsmath,color}
\usepackage{nccmath}

\usepackage[pdfstartview=FitH,colorlinks=true,citecolor=blue,linkcolor=blue,urlcolor=blue]{hyperref}

\begin{document}
\title{Statistical properties of the momentum occupation numbers of the Tonks-Girardeau gas in a harmonic trap}
\author{P. Devillard}
\affiliation{Aix Marseille Univ, Universit\'e de Toulon, CNRS, CPT, Marseille, France}
\author{A. Benzahi}
\affiliation{Universit\'e C\^ote d'Azur, CNRS, Institut de Physique de Nice, France}
\author{P. Vignolo}
\affiliation{Universit\'e C\^ote d'Azur, CNRS, Institut de Physique de Nice, France}
\author{M. Albert}
\affiliation{Universit\'e C\^ote d'Azur, CNRS, Institut de Physique de Nice, France}

\begin{abstract}
  We compute the fluctuations of the number of bosons with a given momentum for the Tonks-Girardeau gas at zero and finite temperature in a harmonic trap. We show that correlations between opposite momentum states $p$, which is an important fingerprint of long range order in  weakly interacting Bose systems are suppressed. Non trivial correlations, including negative correlations are observed for momenta smaller or of the order of the inverse radius of the gas. The full distribution of the number of bosons with momentum $p$ exhibits an interesting crossover from a non trivial distribution at zero momentum to an exponential distribution. The distribution of the quasi-condensate occupation is also studied. Experimental relevance of our findings for recent cold atoms experiments are discussed.
\end{abstract} 

\maketitle

\section{Introduction}
\label{sec_intro}

The Tonks-Girardeau gas is a very peculiar state of matter made of one-dimensional bosons with infinite local repulsion \cite{Girardeau,Giamarchireview,Caza2011}. If it was considered as a toy model for theoretical physicists for a long period of time, it is now an established playground for the study of strongly correlated quantum physics \cite{Parendes2004,Weiss2004,Weiss2005} and is an important system to benchmark results of quantum simulators since it is an exactly solvable model. It shares many common aspects with the gas of free fermions, as far as the observables which depend only on the diagonal elements of the density matrix, such as density-density correlations, are concerned. However, this is no longer the case for quantities which depend on the off-diagonal elements of the density matrix \cite{Caza2011}. One of them is the momentum distribution, namely the average number of bosons carrying a momentum $p$, which is routinely measured in ultra-cold atom experiments. This observable has been used, for instance, to probe Bose-Einstein condensation in weakly interacting quantum gases \cite{Cornell1995,Ketterle1995}, to measure quantum depletion due to interactions \cite{Bogo1947,ClementAspect,Lopes2017}, to observe the superfluid to Mott insulator transition \cite{Greiner2002} and is also very sensitive to interactions \cite{Giamarchireview,Lenard,Leggett2006} and many-body symmetries of the wave function \cite{Pagano2014,Decamp2016}.

Beyond the knowledge of the average values, correlations between different momentum occupation numbers and their fluctuations shed light to many interesting phenomena. For example, perfect correlations between opposite momenta, was predicted by Bogoliubov theory in weakly interacting systems \cite{Bogo1947} and identified to be a hallmark of Bose-Einstein condensation \cite{FangBouchoule,Tenart2021}. Among others, correlations in momentum space have be shown to be useful for analyzing phenomena such as dynamical Casimir effect \cite{Jaskula2012}, Hawking radiation \cite{Unruh1981,Balbinot2008,Recati2009,Fabbri2018,Steinhauer2019} or the escape from a barrier \cite{Dobrz2019}. In this paper, we study the fluctuations of the momentum occupation number $\hat n_p$ in a gas of one-dimensional bosons in the Tonks-Girardeau limit at zero and finite temperature in a harmonic trap. This fills the gap with previous works in the weakly interacting regime \cite{Bouchoule2012,FangBouchoule}, in the hydrodynamic regime (low energy) with arbitrary interaction but in the absence of a trap \cite{Mathey2009,LovasDoraDemlerZarand,LovasDoraDemlerZarand2} or in the Tonks regime at zero temperature on a ring \cite{Rigol2011,Devillard2020}. In addition, we discuss the statistical distribution of the number of particles in the lowest natural orbital of the system (quasi-condensate state) beyond the results on the average value obtained in Refs. \cite{ForresterFrankelGaroniWitte,Papenbrock2003}.

This article is organized as follows. We start by presenting the model in Sec. \ref{sec_model} and explaining the general formalism to compute the correlations. Section \ref{sec_corr} is then devoted to the calculation of the second moment $\langle \hat n_p^2 \rangle$ and the correlations $\langle \hat n_p \hat n_q \rangle$ of the momentum occupation number. In Sec. \ref{sec_fcs}, we determine the moments $\langle \hat n_p^k\rangle$ for all integer $k$ and reconstruct the full counting statistics of $\hat n_p$ (FCS). In Sec. \ref{sec_qc}, we discuss the probability distribution of the fundamental natural orbital. Finite temperature and experimental considerations are discussed in Sec. \ref{sec_exp} and our main conclusions are summarized in Sec. \ref{sec_ccl}. Natural extensions of this work are sketched and we mention some perspectives for future studies. In addition, several technical details are given in appendixes \ref{app_hom_gas}--\ref{app_rho1_T}.


\section{Model} 
\label{sec_model}
We consider a gas of $N$ identical bosons of mass $m$ confined in a one-dimensional harmonic trap of frequency $\omega$ at zero temperature. The Hamiltonian of the system reads
\begin{equation}
\mathcal H \, = \,\sum_{i=1}^{N} \left(-\frac{\hbar^2}{2m} \frac{\partial^2}{\partial x_i^2}+\frac{1}{2}m\omega^2 x_i^2\right)  \, + \, g \sum_{i<j} \delta (x_i - x_j),
\end{equation}
where $x_i$ is the position of the $i^{th}$ bosonic particle, $g=-2\hbar^2/(ma_{1D})$, with $a_{1D}$ the effective one-dimensional scattering length \cite{Olshanii1998}. In this article, we will focus on the Tonks-Girardeau limit where $g$ is sent to infinity. In this regime, the ground state is constructed by filling all single particle orbitals up to the Fermi energy $E_F=\hbar \omega N$ while preserving the bosonic statistics as described below. This is the so called regime of fermionization where all physical observables which depend only on density or density correlations are similar to the ones of a perfect gas of fermions \cite{Girardeau,Korepinetal}. At zero temperature, the many-body wave function of the gas is given by

\begin{equation}
\label{eq_psi}
  \psi(\{x_i\})=\mathcal N\prod_{k=1}^{N}e^{-(x_k/a_0)^2/2}\prod_{1\le j\le k\le N}|x_j-x_k|,
\end{equation}
with $a_0=\sqrt{\hbar/m\omega}$ the oscillator length and $\mathcal N=1/\sqrt{a_0^N N!\prod_{m=0}^{N-1} 2^{-m}\sqrt{\pi}m!}$. In the large $N$ limit, the density profile takes the form of a semi-circle for both bosons and fermions

\begin{equation}
  \label{eq_density_trap}
   n(x)=\sqrt{\frac{2N}{(\pi a_0)^2}}
\sqrt{1-\Bigl(\frac{x}{\sqrt{2Na_0^2}}\Bigr)^2},
\end{equation}
and defines the radius of the cloud $R=\sqrt{2N}a_0$, which will be a very important parameter in this study.

Quantum statistics enters into play whenever off-diagonal elements of the density matrix are involved in an observable. The one-particle density matrix itself, defined as

\begin{equation}
  \label{eq_md}
  \rho_1(x,x')=\langle\hat \Psi^\dagger(x)\hat\Psi(x')\rangle=\int  \psi^*(X)\,\psi (X')\, dx_2\cdots dx_N,
\end{equation}
where $\hat \Psi(x)$ is the bosonic field operator, $X=(x,x_2,...,x_N)$, $X'=(x',x_2,...,x_N)$, is indeed very sensitive to quantum statistics and quantum fluctuations. It has been used for instance to construct the phase diagram of a one-dimensional gas in a harmonic trap \cite{Petrov2000}. In the Tonks-Girardeau regime, its off diagonal part decays algebraically as $|x-x'|^{-1/2}$ and therefore prohibits Bose-Einstein condensation since long range order is not possible in the thermodynamic limit. Its Fourier transform, which gives the momentum distribution, is a commonly used observable, that gives the average number of bosons with a momentum $p$

\begin{equation}
  \label{eq_md2}
  \langle \hat n_p\rangle =\langle \hat a^\dagger_p \hat a_p\rangle=\iint  e^{-i p (x-x')/\hbar}\rho_1(x,x')\,dx \, dx',
\end{equation}
where $\hat a_p=\int dx\,e^{i p x/\hbar} \hat \Psi(x)$ is the annihilation operator of a particle with momentum $p$. In a harmonic trap, it has been calculated \cite{Papenbrock2003,MinguzziVignoloTosi,Rigol2015} and measured \cite{Parendes2004,Wilson2020} and shown to display the following features. The peak at small momenta shrinks due to interactions and a tail develops (this is actually true for any value of the interaction parameter and not only in the Tonks-Girardeau limit). This tail decays algebraically as $C p^{-4}$, with $C$ the Tan's contact \cite{Tantheoretical}, for momenta larger than the Fermi momentum $p_F=\hbar\pi n(0)$. In this paper, we will treat these two regimes differently and refer to them as hydrodynamic regime ($p<p_F$) and Tan regime ($p>p_F$).

\section{Correlations in momentum space} 
\label{sec_corr}

As discussed before, the momentum distribution provides important information about quantum statistics and quantum fluctuations. However, it does not provide precise information about the correlations between particles which is the subject of this article. We therefore turn now to the description of the fluctuations of $\hat n_p$, the number operator of particles with momentum $p$, and the correlations between different momenta. We then define the following quantity

\begin{equation}
  \mathcal G_{p,q}=\langle \hat n_p \hat n_q\rangle/\langle \hat n_p\rangle \langle\hat n_q\rangle-1.
\end{equation}
Its diagonal part is simply the noise to signal ratio square while the off-diagonal part describes correlations between different momentum occupation numbers. It gives information about the joint probability to detect atoms with momentum $p$ and $q$. If these atoms are not correlated, this quantity is simply zero.

In order to calculate the fluctuations of $n_p$, we shall need the two-body density matrix, defined as
\begin{eqnarray}\label{eq_rho2_def}
  \rho_2(x,u;y,w)=\langle\hat \Psi^\dagger(x) \hat \Psi^\dagger(u) \hat \Psi(y) \hat \Psi(w)\rangle.
\end{eqnarray}
Using the definition of the number operator $\hat n_p$ given before and standard bosonic commutation relations, we have
\begin{eqnarray}\label{eq_Npq}
  \langle \hat n_p \hat n_q \rangle  &=&  \int e^{i\frac{p(y-x)}{\hbar}} e^{i\frac{q (w-u)}{\hbar}}\rho_2 \,dx \, dy \, du \, dw \nonumber\\
  & &+\;\delta_{p,q} \langle \hat n_p\rangle .
\end{eqnarray}
The last term in (\ref{eq_Npq}) is known as the shot noise term. In the large $N$ limit, it is negligible compared to the first contribution and will be omitted in this work.

\subsection{Hydrodynamic regime}

We start the discussion with the low momentum regime. In that case, the one-dimensional Bose gas can be described by a low energy theory known as the harmonic fluid approach \cite{Haldane} or Luttinger liquid theory \cite{Giamarchireview}. In the presence of a harmonic trap, it is possible to compute the one-body density matrix at zero temperature with various methods \cite{Petrov2000,Gangardt2003,Mora2003,ForresterFrankelGaroniWitte,Papenbrock2003,Gangardt,Dubail2017} which reads

\begin{equation}
  \label{eq_rho1_trap}
  \rho_1(x,y)=\frac{G(3/2)}{\sqrt{2\pi}}\frac{ n(x)^\frac{1}{4} n(y)^\frac{1}{4}}{|x-y|^\frac{1}{2}},
\end{equation}
where the average density $n(x)=\langle\hat \Psi^\dagger(x)\hat \Psi(x)\rangle$ is given by the semi-circle law (\ref{eq_density_trap}) and $G$ is the Barnes function \cite{Grasd}. However, the precise value of the constant is not needed for the calculation of $\mathcal G_{p,q}$. Surprisingly, the form of the density matrix is exactly the one of a uniform system with the replacement of the density by a local density. The power law decay of correlations is the same for instance. Of course, this formula will be valid only for long distances or small momentum, this is why this section will be restricted to momenta smaller than the Fermi momentum.

Since the harmonic fluid theory is Gaussian, it is then possible to compute the higher order density matrix, using Wick's theorem which in this case reads \cite{Tsvelik1998}
\begin{widetext}
\begin{align}
  \langle \hat\Psi^\dagger(x_1)\cdots \hat\Psi^\dagger(x_n)\,\hat\Psi(x'_1)\cdots \hat\Psi(x'_n)\rangle=\frac{\prod_{i,j} \langle \hat\Psi^\dagger(x_i) \hat\Psi(x'_j)\rangle}{\prod_{i<j} \langle \hat\Psi^\dagger(x_i) \hat\Psi(x_j)\rangle \prod_{i<j} \langle \hat\Psi^\dagger(x'_i) \hat\Psi(x'_j)\rangle},
  \label{eq_wick}
\end{align}
\end{widetext}
with $\hat\Psi(x_i)$ is the bosonic field operator. We compute numerically the average number and the variance of the number of bosons with a given momentum $p$ with this prescription using a Metropolis algorithm. The average $\langle n_p\rangle$ is found to be in agreement with already known results \cite{ForresterFrankelGaroniWitte,Papenbrock2003}.

\begin{figure*} 
  \includegraphics[width=0.5\linewidth]{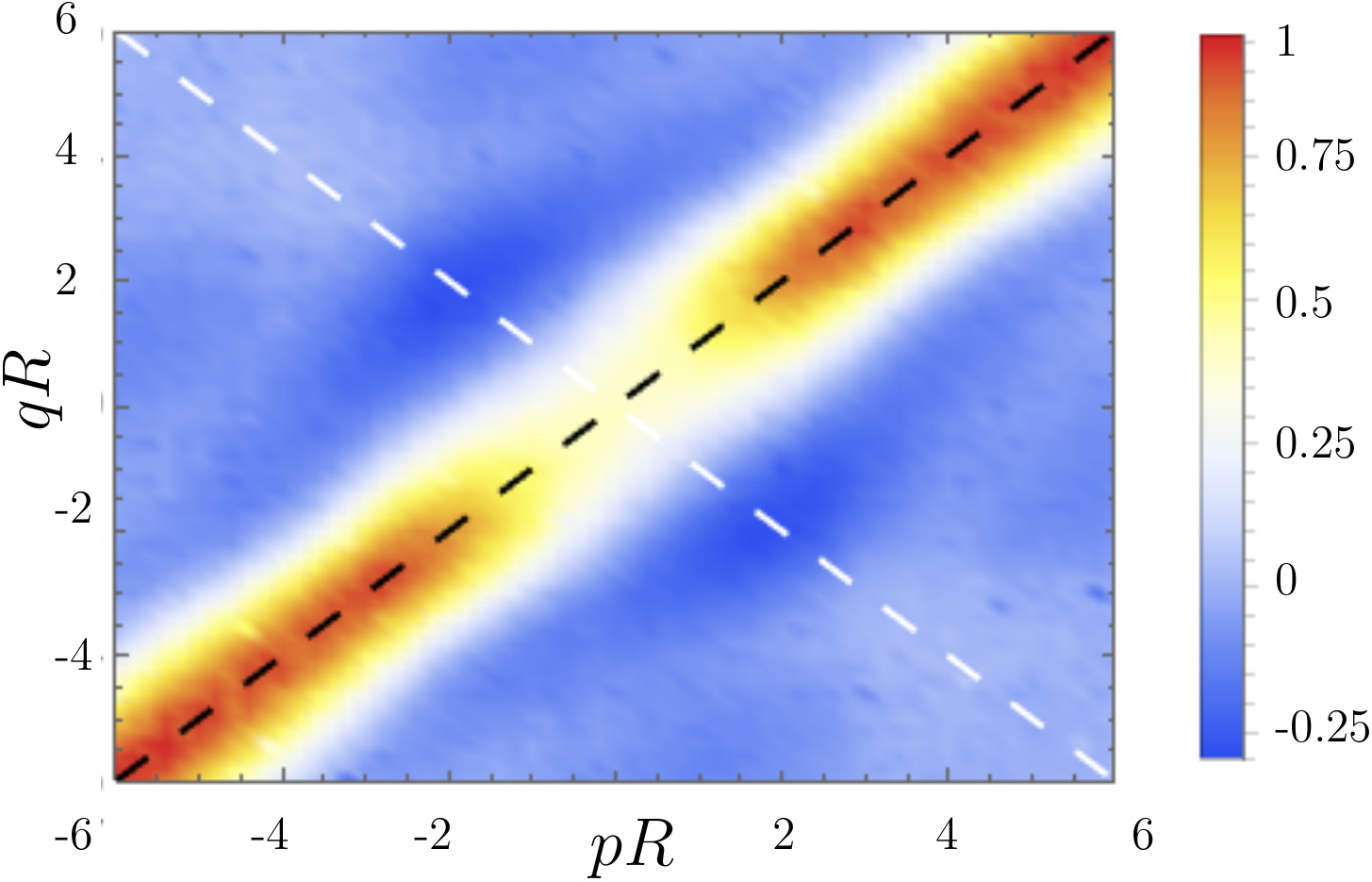}
  \hspace{0.01\linewidth}
  \includegraphics[width=0.46\linewidth]{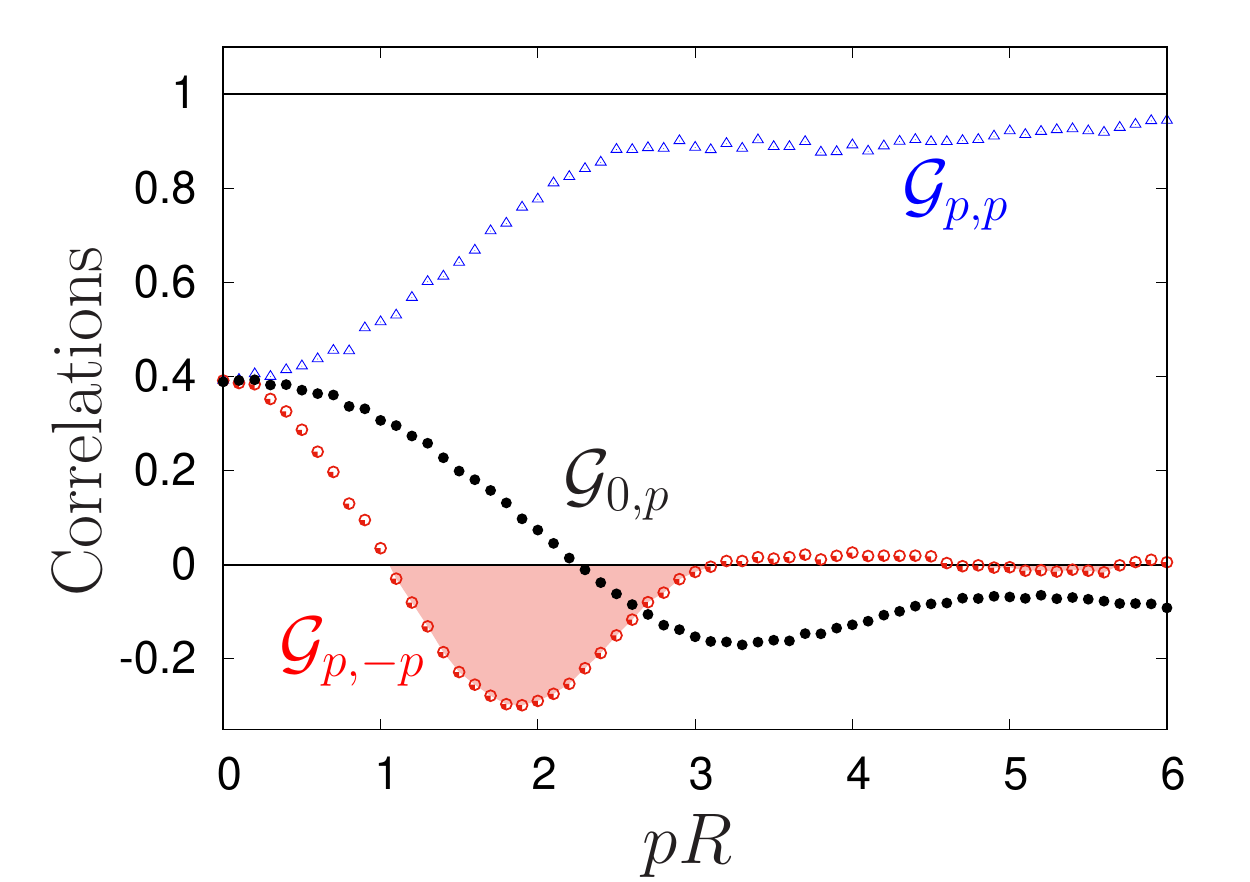}
  \caption{\label{fig_corr} Left: density plot of the second order correlations $\mathcal G_{q,p}$= $\langle \hat n_p \hat n_q\rangle/\langle \hat n_p\rangle\langle \hat n_q\rangle-1$ of the Tonks gas at zero temperature in a harmonic trap. Momenta $p$ and $q$ are expressed in terms of the inverse of the radius $R$ of the gas. Dashed lines are guides to the eye and correspond to equal momenta ($p=q$, black) and opposite momenta ($p=-q$, white). Right: Cut of the momentum occupation number correlations $\mathcal G_{q,p}$ along the lines $p=q$ (blue triangles), $q=0$ (black dots) and $p=-q$ (red circles). The red area emphasizes the region where $\mathcal G_{p,-p}$ is negative.}
\end{figure*}

Figure \ref{fig_corr} shows the results for $\mathcal G_{p,q}$ as a function of $p$ and $q$. Correlations between occupation numbers are mainly present on the diagonal $p=q$ as already observed in the absence of a trap \cite{Mathey2009,Rigol2011,Devillard2020}. In particular, along the anti-diagonal ($p=-q$), perfect correlations between atoms that are typical of Bogoliubov excitations in the weakly interacting regime are strongly suppressed as already discussed in previous works \cite{Devillard2020, Bouchoule2012,FangBouchoule}. The precise behavior of $\mathcal G_{p,q}$ is shown on Fig. \ref{fig_corr} with cuts of this function along the diagonal, the anti-diagonal ($p=-q$) and correlations with the zero momentum state. Similar results for homogeneous systems in a ring and a box geometry can be found in Appendix \ref{app_hom_gas} for comparison. At very small momentum, the statistics of $\hat n_p$ is clearly sub-exponential with $\mathcal G_{p,p}<1$ ($\langle \hat n^2_p\rangle<2\langle \hat n_p\rangle^2$) and tends to an exponential statistics for $pR>6$ which corresponds to the standard bunching effect of non-interacting bosonic particles. This will be corroborated by the study of the FCS of $\hat n_p$ in the next section. 

Along the anti-diagonal, anti-correlations for $\mathcal G_{p,-p}$ are visible for the trap in the range of $p$ between $1$ and $3$ times $1/R$. Such correlations are totally absent in the case of interacting bosons on a ring \cite{Mathey2009,Devillard2020} but also exist in a box geometry as discussed in Appendix \ref{app_hom_gas}. They have also been observed in the weakly interacting regime in a box geometry \cite{Bouchoule2012} and in a harmonic trap \cite{FangBouchoule}. Moreover, we will see in the next section that this effect is destroyed by temperature, which leads us to assume that these negative correlations are due to an interference effect. This is confirmed by an analytical argument given in Appendix \ref{app_hom_gas}. 

\subsection{High momentum regime}

In this regime, bosonization is no longer applicable and we have to resort to a more sophisticated small distance expansion of the density matrix known as Lenard's expansion \cite{Lenard}. Along the lines of Ref. \cite{Devillard2020}, we find that the occupation numbers are uncorrelated and obey an exponential distribution. Lenard's expansion expresses the $n$-body bosonic density matrix in terms of the fermionic one. We give here the expansion of the $2$-body density matrix for the trap at zero temperature; the expansion for the $n$-body density matrix in the general case of finite temperature is given in appendix \ref{app_lenard}. The bosonic one-particle density matrix reads

\begin{eqnarray}\label{eq_lenard}
  \rho_B(x,u;y,w) \, & = & \, sgn(u-x) \, sgn(w-y) 
  \Bigl\lbrack
  \langle x,u \vert \rho_F \vert y,w \rangle \, + \, \nonumber \\
  (-2) &\, & \int_J \langle x,u,x_3 \vert \rho_F \vert y,w,x_3 \rangle dx_3 + ... \nonumber \\
  + \, {(-2)^n \over n!} \int_J ... & \, & \int_J  \langle x,u,x_3,...,x_n \vert \rho_F \vert y,w,x_3,...,x_n \rangle
  \Bigr\rbrack \nonumber \\
  &\, & \,\,\,\,\,\,\,\,\,\,\,\,\,\,\,\,\,\,\,\,\,\, dx_3 \, ... dx_n, 
\end{eqnarray}
where $J$ is the interval $\lbrack x,y\rbrack  \cup \lbrack u,w\rbrack$ and $sgn$ is the sign function. The $n$-body fermionic density matrix reads \cite{MinguzziVignoloTosi,Minguzzi2013}
\begin{eqnarray}
  &\,&\langle x,u,x_3,...,x_n \vert \rho_F \vert y,w,x_3,...,x_n \rangle \, = 
  \nonumber\, \\
  &\,&\left\vert
  \begin{matrix}
    K(y,x) & K(w,x) & K(x_3,x) & ...& ...& K(x_n,x) \cr
    K(y,u) & K(w,u) & K(x_3,u) & ...& ...& K(x_n,u) \cr
    K(y,x_3) & K(w,x_3) & \sqrt{2N} & ...& ...& K(x_n,x_3)\cr
    ...    & ...    & ...      & ...& ...& ...    \cr
    K(y,x_n) & K(w,x_n)& K(x_3,x_n) & ...& ...& \sqrt{2N}
  \end{matrix}
  \right\vert, \nonumber \\
\end{eqnarray}
with the kernel
\begin{equation}
  K(x,y) \, = \, \sum_{\nu_1, ...,\nu_N} \Theta(-E_{\nu}) \sum_{l=1}^N 
  u_{\nu_l}(x) u_{\nu_l}(y),
\end{equation}
$E_{\nu} = \sum_{l=1}^N \epsilon_{\nu_l}$ and $\Theta$ is the Heaviside function; $u_{\nu_l}$ are the eigenfunctions of the harmonic oscillator with energy level  $\epsilon_{\nu_l} = (\nu_l + 1/2) \hbar \omega$. The first sum runs over all possible sequences $(\nu_1,\nu_2,...,\nu_N)$. In the thermodynamic limit and to lowest order in $1/p$, each determinant giving a relevant contribution to the sum factorizes into dipoles. For example, the term for $n=2$, $\langle x,u,x_3,x_4 \vert \rho_F \vert y,w,x_3,x_4 \rangle$ reduces to $\langle x,x_3 \vert \rho_F \vert y, x_3 \rangle \,  \langle u,x_4 \vert \rho_F \vert w, x_4 \rangle  + ( y \leftrightarrow w)$. When computing $\langle n_p^2\rangle$, the main contribution will come from two configurations of dipoles; the direct term corresponds to $x$ close to $y$ and $u$ close to $w$ and the exchange term, which is exactly the reverse, $x$ is close to $w$ and $y$ is close to $u$. For $\langle n_p^2 \rangle$ (but not for $\langle n_p n_q\rangle$ if $q \not= p$), the two terms are equal and will give an overall contribution to $\langle n_p^2 \rangle$ which behaves as $1/p^8$, which finally implies $\langle n_p^2 \rangle = 2 \langle n_p\rangle^2$ and $\mathcal G_{p,p} \simeq 1$.  
   
Repeating the argument for $\langle n_p^n\rangle$, we obtain $\langle n_p^n \rangle \, = \, n! \langle n_p \rangle^n$ and thus, the probability $P(n_p)$ is exponential. The argument can be carried out also for the calculation of the correlations $\langle n_p n_q \rangle$ but the exchange term is very small and to lowest order, there are no correlations between $n_p$ and $n_q$ for $q \not= p$. So far, we have focused on the main term. By considering higher order terms, we found that correlations between $n_p$ and $n_q$ do exist. A rough estimate of the correlation $\mathcal G_{p,q}$, in Tan contact's regime yields an order of magnitude $p^{-4}q^{-4} (p^{-1} + q^{-1})$, which is extremely small and out of reach of current experimental capabilities. Technical details can be found in Appendix \ref{app_tan}.

\section{Full counting statistics of the momentum occupation number} 
\label{sec_fcs}

\begin{figure} 
  \includegraphics[width=\linewidth]{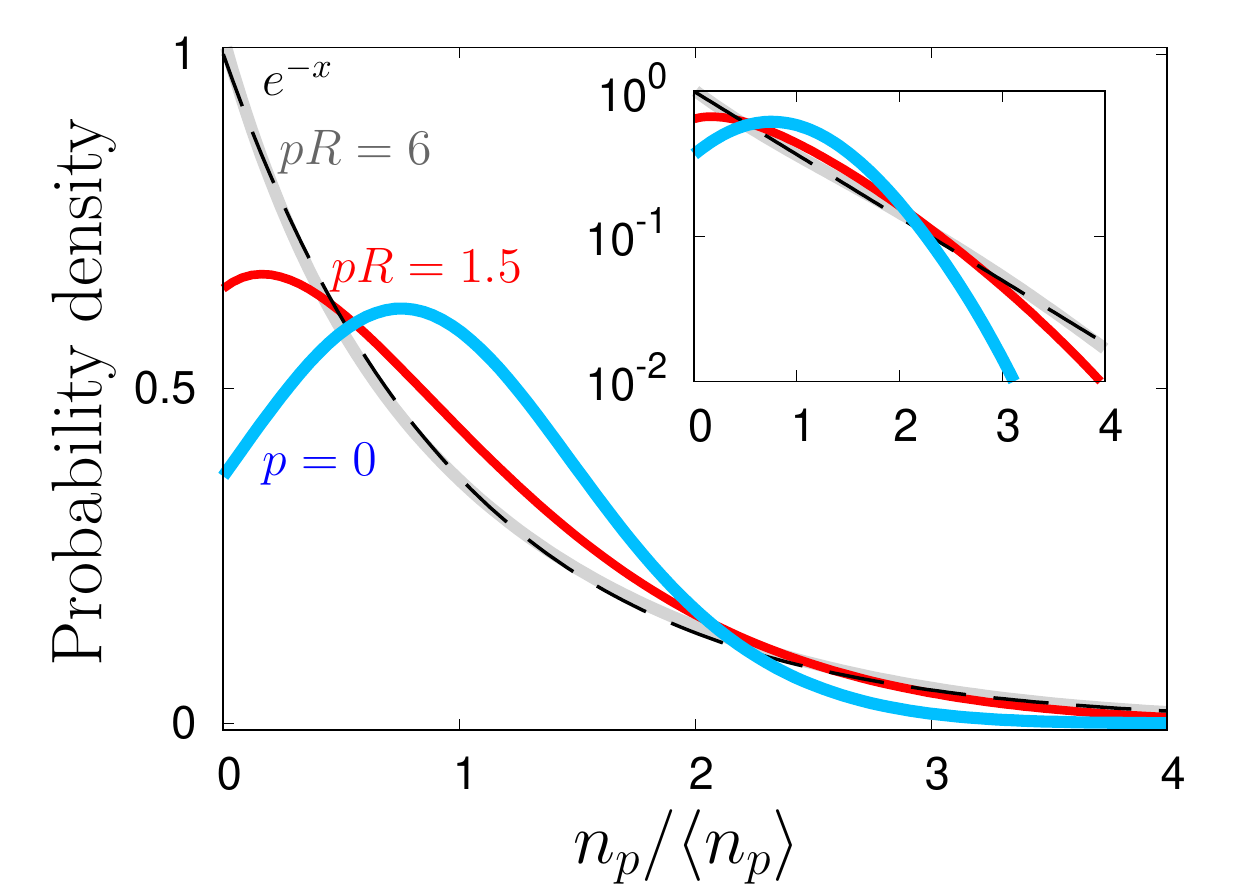}
  \caption{\label{fig_pnp} Probability densities of $n_p/\langle n_p\rangle$ for different values of the momentum $p$ (expressed in terms of the inverse radius of the gas). As the momentum is increased, the distribution tends to an exponential (dashed black line). The inset shows the same data in logarithmic scale.}
\end{figure}

We now turn to the discussion of the FCS of the occupation numbers $n_p$. Using bosonization, we can compute numerically the moments of $n_p$ to arbitrary order and reconstruct the distribution from them as explained in appendix \ref{app_proba_reconstruct}. As in the case of the homogeneous Tonks gas \cite{Devillard2020}, we observe a crossover from a non-trivial distribution \cite{AltmanDemlerLukin,LovasDoraDemlerZarand,LovasDoraDemlerZarand2,Devillard2020} close to zero momentum to an exponential distribution at intermediate and large momentum. The results are presented on Fig. \ref{fig_pnp}. Strong correlations between low momentum states appear one more time to affect the occupation number distribution whereas at larger momentum, correlations vanish and the distribution is exponential.

Concerning the zero momentum state, an important distinction with the homogeneous case has to be made. In the trapped case, $\langle n_0 \rangle$ is proportional to $N$ \cite{Papenbrock2003} and not to $\sqrt{N}$ \cite{Lenard,ForresterFrankelGaroniWitte}. This is related to the fact that any even orbital of the harmonic oscillator has a non-zero overlap with the $p=0$ state. However, this has no relation to Bose-Einstein condensation. The quasi-condensate mode is defined in that case as the eigenstate of the one-body density matrix with the largest occupation number and is very different from the zero momentum state. This point will be discussed in the next section.

For the zero momentum occupation number, we obtain the following numerical values for $\mathcal F_k=\frac{\langle n_0^k\rangle}{\langle n_0\rangle^k}$,
$$
\begin{array}{|c|c|c|c|c|c|c|}
  \hline
  \mathcal F_2 & \mathcal F_3 & \mathcal F_4 &\mathcal F_5 & \mathcal F_6 & \mathcal F_7 & \mathcal F_8 \\
  \hline
  1.395 & 2.356 & 4.566 & 9.849 & 23.167& 58.806 & 159.198\\
  \hline
\end{array}
$$
with $\langle n_0\rangle\simeq 0.62\, N$ as already calculated in Refs. \cite{ForresterFrankelGaroniWitte,Papenbrock2003}. From this finite number of moments, we reconstruct the distribution with a fairly good accuracy. Indeed, we have checked the accuracy by using different reconstruction algorithms (see Appendix \ref{app_proba_reconstruct}) and by comparing the results with known distributions such as the one for large momentum (exponential) or the distribution of zero momentum in the ring geometry where it has been calculated exactly \cite{Devillard2020,Gritsevetal1,Gritsevetal2}. 

The result is presented on Fig. \ref{fig_pnp} along with the distribution of finite momentum occupation numbers. The precise shape of the distribution is not quite different from the homogeneous case, as shown on Fig. \ref{fig_pqc} except that the value at zero is smaller, meaning that the probability to find zero particles with $p=0$ is reduced. However, we insist on the fact that the moments do not have the same scaling with the number of particles $N$. The transition to an exponential distribution occurs for momenta larger than a few times the inverse of the cloud radius. This is expected since short wavelength excitations are not very sensitive to boundary effects and therefore the homogeneous result is recovered \cite{Devillard2020}. Finally, at large momenta where bosonization is not applicable, we have shown in the previous section that the distribution was exponential.

\section{Quasi-condensate mode}
\label{sec_qc}

Although rather academical, we now raise the question of the statistical properties of the occupation of the quasi-condensate mode. For a Tonks-Girardeau gas in a harmonic trap, the quasi-condensate mode is no longer the state with zero momentum nor the single particle ground state but a rather complicated state which depends on interactions. Following the so-called Penrose-Onsager criterion for Bose-Einstein condensation in interacting systems, it is defined as the eigenstate of the one-body density matrix with the largest occupation number (see \cite{ForresterFrankelGaroniWitte,Papenbrock2003} for instance). The eigenmodes $\phi_j$ of the one-body density matrix with eigenvalue $\mu_j$ are defined as

\begin{equation}\label{eq_qc}
  \int  \rho_1(x,y)\phi_j(y)\, dy \, = \, \mu_j \phi_j(x).
\end{equation}
Of course, there is no Bose-Einstein condensate in the Tonks-Girardeau regime, since the largest eigenvalue of the density matrix is not proportional to the number of particle and is not isolated from the other modes. Indeed, it is well known in this regime that the average occupation of these modes scales with $\sqrt{N}$ and decays algebraically with $j$ \cite{ForresterFrankelGaroniWitte,Papenbrock2003}. Therefore, there is no macroscopically occupied single mode, namely no Bose-Einstein condensation. Still, we discuss here the FCS of the quasi-condensate at zero temperature for curiosity.

Once the eigenmodes of the one-particle density matrix are known, it is a simple quantum mechanics exercise to compute the moments of the occupation numbers $\hat n_{\phi_j}=\hat a^\dagger_{\phi_j}\hat a_{\phi_j}$, with

\begin{equation}
   \hat a_{\phi_j}=\int  \phi_j(x)\hat \Psi(x) \, dx,
\end{equation}
following the same method as before. In particular, for the quasi-condensate mode, also called the fundamental orbital, we can use bosonization to compute the correlations functions of $\hat \Psi$ and $\hat \Psi^\dagger$. We solve Eq. (\ref{eq_qc}) by discretization and numerical diagonalization of the one-particle density matrix. The occupation number of the fundamental orbital shall be denoted by $n_{\phi_0}$. This yields the following results for the rescaled moments $\mathcal Q_k=\langle \hat n^k_{\phi_0}\rangle/\langle \hat n_{\phi_0}\rangle^k$

\begin{figure} 
  \includegraphics[width=\linewidth]{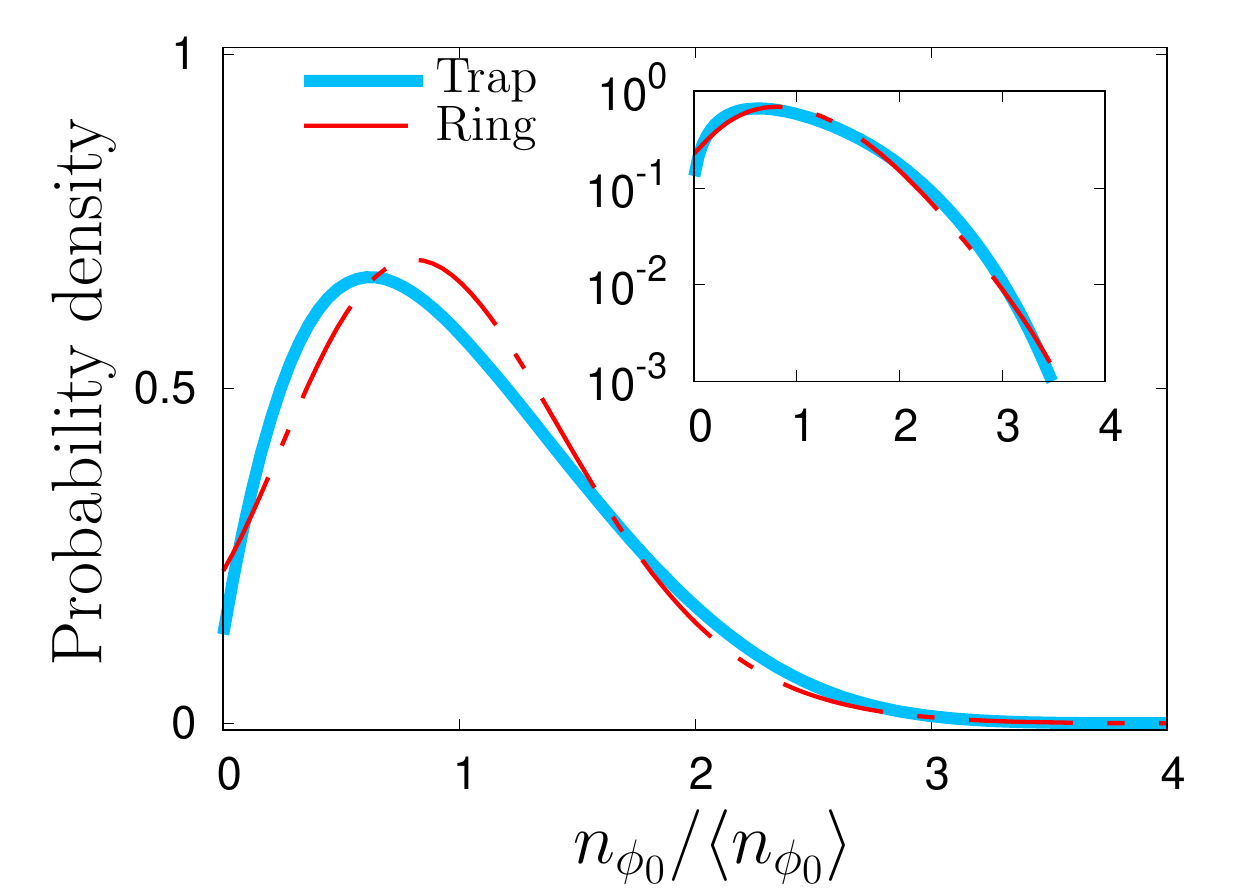}
  \caption{\label{fig_pqc} Probability densities of the quasi-condensate occupation number $n_{\phi_0}/\langle n_{\phi_0}\rangle$ in the harmonic trap (thick blue line) and in the a ring geometry (red dashed line from Ref. \cite{Devillard2020}) in the Tonks-Girardeau regime at zero temperature. The inset shows the same data in logarithmic scale.}
\end{figure}

$$
\begin{array}{|c|c|c|c|c|c|c|}
  \hline
  \mathcal Q_2 & \mathcal Q_3 & \mathcal Q_4 &\mathcal Q_5 & \mathcal Q_6 & \mathcal Q_7 & \mathcal Q_8 \\
  \hline
  1.372 & 2.279 & 4.333 & 9.159 & 21.129 & 52.75 & 139.94 \\
  \hline
\end{array}
$$
with $\langle \hat n_{\phi_0}\rangle=3.438\sqrt{N}$ (in perfect agreement with Ref. \cite{ForresterFrankelGaroniWitte}). We note in particular that $\langle \hat n^2_{\phi_0}\rangle=1.372\langle \hat n_{\phi_0}\rangle^2$, which is below the prediction for an exponential distribution by a factor almost two. Again, we reconstruct the full probability distribution from the available moments and obtain the results of Fig. \ref{fig_pqc}. We compare our result to the one we obtained in \cite{Devillard2020} in a ring geometry. In that case, the quasi-condensate mode is the zero momentum state. Once rescaled by their repective mean values, the distribution of $n_{\phi_0}/\langle n_{\phi_0}\rangle$ are rather similar. There is however a small tendency for condensation in a trap geometry, since the probability to observe zero particle in the quasi-condensate is lower in that case than for a ring geometry.

\section{Experimental considerations} 
\label{sec_exp}

Up to now, we have considered the zero temperature physics of a Tonks-Girardeau gas in a harmonic trap in perfect conditions. In this section, we discuss some experimental issues for the detection of momentum occupation number correlations such as the effects of a finite temperature, the shot noise and the imperfection of the detection process. However, we emphasize that such measurements of correlations in momentum space are now standard for several experimental groups and have been done in the weakly interacting regime in one dimension \cite{FangBouchoule} and for arbitrary interaction in three dimensions \cite{Tenart2021,Carcy2019}.

\subsection{Effect of finite temperature}

According to the phase diagram of the one-dimensional gas in a harmonic trap \cite{Petrov2000}, increasing the temperature will destroy the Tonks-Girardeau state and bring the system into a classical gas phase. This will be governed by phase fluctuations which lead to an exponential decay of the off-diagonal part of the one-body density matrix instead of an algebraic one. As a consequence, momentum correlations will be strongly affected by finite temperature.  

We start by discussing the low momentum regime. In that case, we explain in Appendix \ref{app_rho1_T} that the one-particle density matrix develops an exponential decay over a length scale $\ell_T$, the thermal length (not to be confused with the de Broglie wave length), which is proportional to the Fermi velocity in the center of the trap divided by the temperature. The crucial parameter is therefore, the ratio of this length with respect to the cloud radius $R$. We show that it is given by

\begin{equation}\label{eq_LT}
  \ell_T/R=\frac{\hbar\omega}{k_B T}=\frac{1}{N} \frac{E_F}{k_B T},
\end{equation}
with $E_F=N\hbar\omega$, the Fermi energy of the gas. In addition, the density profile is also modified by temperature, since particles are now allowed to occupy higher energy orbitals than the ones below Fermi energy. Again, all the details are given in Appendix \ref{app_rho1_T}.

\begin{figure} 
  \includegraphics[width=0.9\linewidth]{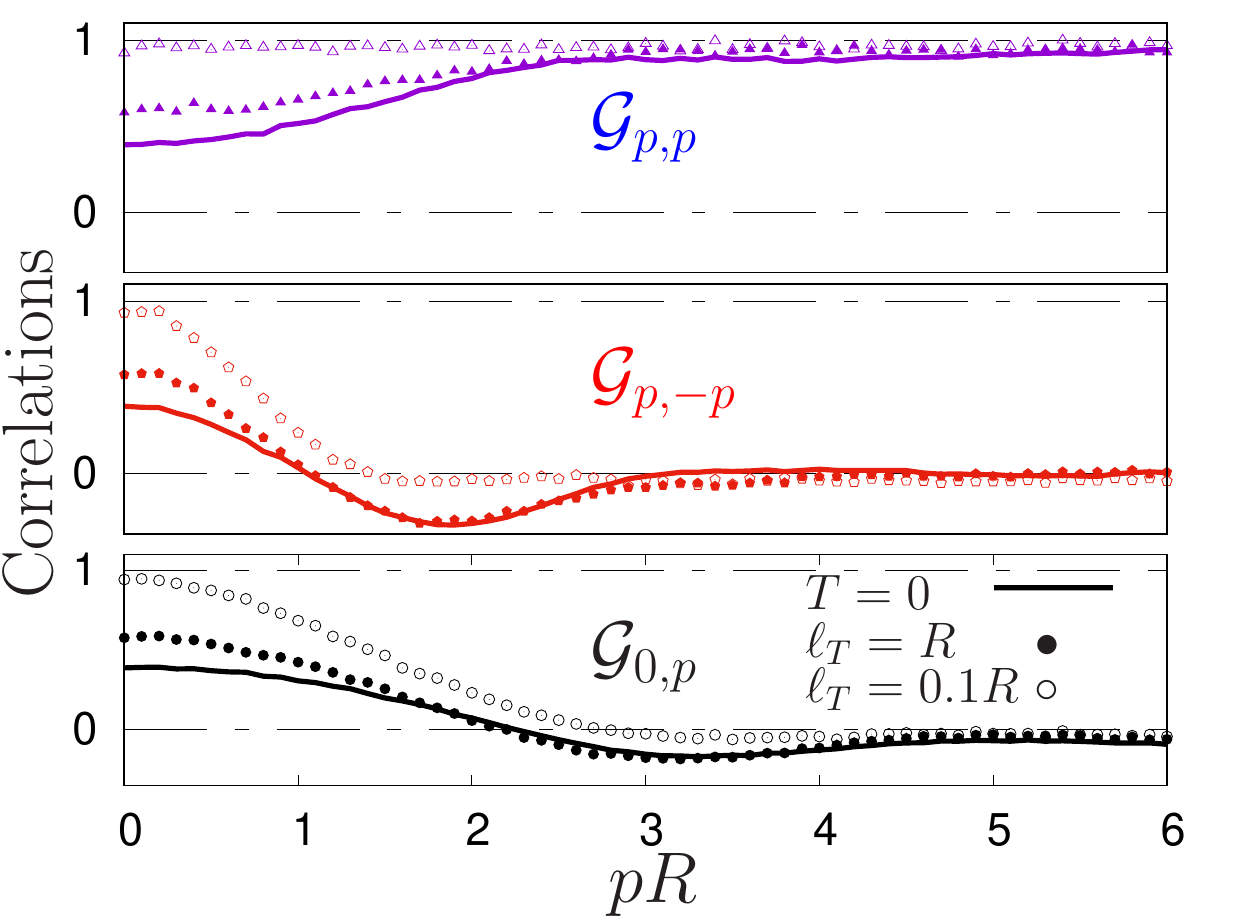}
  \caption{\label{fig_corr_ft} Cut of the momentum occupation number correlations $\mathcal G_{q,p}=\langle \hat n_p \hat n_q\rangle/\langle \hat n_p\rangle\langle \hat n_q\rangle-1$ at finite temperature along the lines $p=q$ (upper panel), $p=-q$ (middle panel) and $q=0$ (lower panel). The full lines are the zero temperature results of Fig. \ref{fig_corr}. Symbols are calculated at temperature $T=0.1 T_F$, with $k_BT_F=N\hbar\omega$, the Fermi energy. Filled symbols correspond to a system with a thermal length equal to the radius of the gas $\ell_T= R$ whereas empty symbols to $\ell_T=0.1 R$.}
\end{figure}

The main conclusion is that, as long as $k_BT \ll E_F$ (typically by a factor ten), the density profile is not really affected. However, the momentum correlations are governed by $\ell_T/R$. If $\ell_T>R$, the zero temperature result is preserved. But as soon as $\ell_T$ becomes smaller than the radius of the cloud, the situation is drastically modified. The results  for $\ell_T=R$ and $\ell_T=0.1\,R$ at $k_BT=0.1 E_F$ ($\ell_T$ is monitored by the number of bosons $N$) are displayed on Fig. \ref{fig_corr_ft}. In the latter case ($\ell_T=0.1\,R$), all the features discussed so far are washed out, in particular the negative correlations in $\mathcal G_{p,-p}$. The diagonal part $\mathcal G_{p,-p}$ quickly reaches the classical prediction of bunching of non-interacting bosons ($\mathcal G_{p,p}=1$) and correlations between different momenta vanish. This is what is expected for a classical gas.

In the high momentum regime, a finite temperature Lenard's expansion shows that dipole correlations are lost; for large $p$, $\mathcal G_{p,p} \simeq 1$ and $\mathcal G_{p,-p} \simeq 0$. Technical details are given in Appendix \ref{app_lenard}.  

\begin{figure} 
  \includegraphics[width=\linewidth]{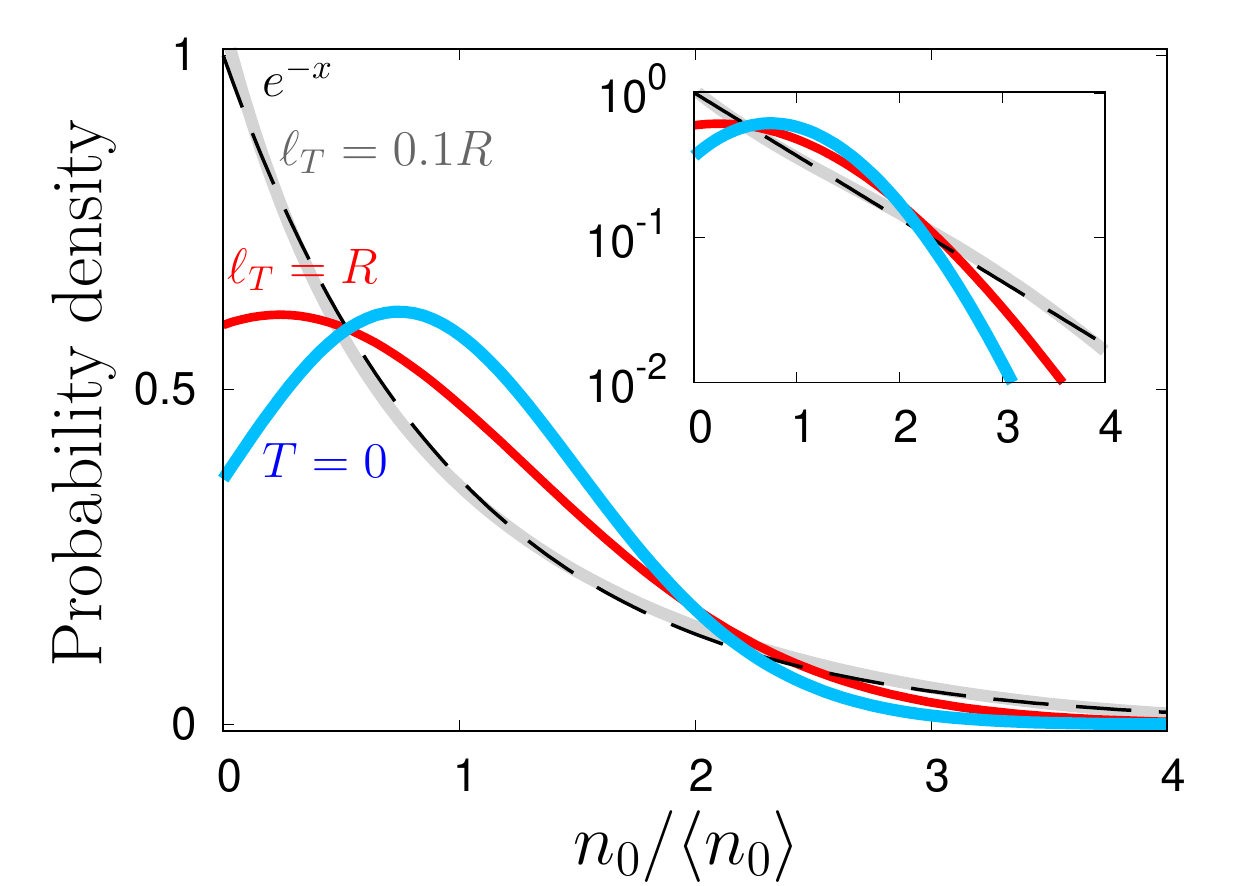}
  \caption{\label{fig_pn0_ft} Probability densities of $n_0/\langle n_0\rangle$ for different thermal lengths $\ell_T$. The temperature is fixed to $T=0.1T_F$ with $T_F=N\hbar\omega/k_B$ the Fermi temperature, $\ell_T=R$ for the red curve and $\ell_T=0.1R$ for the gray curve. As the thermal length is decreased, the distribution tends to an exponential (dashed black line). The inset shows the same data in logarithmic scale.}
\end{figure}

Concerning the probability distribution $P(n_p)$, we expect all of them to converge to an exponential distribution in the large temperature limit. Since this is already the case for momenta larger than $6/R$, we discuss here the most interesting one which is the zero momentum number distribution. We have calculated the moments of $\hat n_0$ at finite temperature with the help of the finite temperature version of the one-body density matrix. Figure \ref{fig_pn0_ft} shows the results for $\ell_T=R$ and $\ell_T=0.1R$ at $k_BT=0.1E_F$. Clearly, the distribution evolves rapidly to an exponential distribution.

In conclusion, in order to observe non-trivial correlations between momentum occupation numbers, it is mandatory to keep the thermal length of the order of the radius of the cloud and to keep $k_BT\ll E_F$. This means that $k_BT$ must be smaller or of the order of $\hbar\omega$ which might be difficult to reach in current experiments.

\subsection{Shot noise and detector efficiency}

From the beginning, we have neglected the shot noise contribution in (\ref{eq_Npq}) arguing that, in the large $N$ limit, its contribution is negligible. This is supported by the fact that $\langle \hat n^k_p\rangle\sim N^k$, therefore its contribution is of order $1/N$ with respect to the main contribution coming from $\rho_k$. However it is very simple to include its contribution in the final result. First, it only affects the diagonal part of $\mathcal G_{p,q}$ and since $\langle \hat n_p\rangle$ is known for the Tonks-Girardeau gas (see \cite{MinguzziVignoloTosi} for instance) it can be easily added if necessary (in the small $N$ regime for instance). Concerning the FCS, this correction might be important only in the large momentum limit where the value of $\langle \hat n_p\rangle$ can be small and therefore $\langle \hat n_p\rangle$ could possibly dominate $\langle \hat n_p\rangle^2$. In that case, it is sufficient to replace the exponential distribution of $n_p$ with a Bose-Einstein distribution

\begin{equation}
  p(n_p)=\frac{1}{(\langle \hat n_p\rangle+1)}\left(\frac{\langle \hat n_p\rangle}{\langle \hat n_p\rangle+1}\right)^{n_p}.
\end{equation}

One remaining important problem is the detection efficiency. Even if single atom detection is improving fastly nowadays, typical experiments in the continuum can detect single atoms with an efficiency at most of the order of $50\%$. In that respect, taking the histogram of $n_p$ over many runs will not correspond to the predicted distribution. Such a low efficiency will convolute the quantum noise and lead to an exponential distribution. To solve this problem, it is customary to measure the normalized moments $\mathcal G_{p,q}$ and $\langle n_p^k\rangle/\langle n_p\rangle^k$ in order to get rid of the small efficiency in the spirit of the measurement of coherence function in quantum optics.


\section{Conclusion and perspectives}
\label{sec_ccl}
In this work, we have proposed a scheme to compute the quantum fluctuations, at zero and finite temperature, of the number of particles $n_p$ with momentum $p$, for the Tonks-Girardeau gas in a harmonic trap.

The correlations between occupation numbers $\mathcal G_{p,q} = \langle \hat n_p \hat n_q \rangle/ \langle \hat n_p\rangle \langle \hat n_q \rangle -1$ are always positive for $p=q$ and show standard bunching for $p$ larger than a few $1/R$. On the contrary, momentum occupation numbers with smaller $p$ show non-trivial correlations. For opposite momentum states, $q=-p$, they are positive for very small momentum but become negative in a range of $p$ between $R^{-1}$ and $3 R^{-1}$. This is at variance with an homogeneous gas on a ring, where all correlations $G_{p,-p}$ are always positive. This phenomenon stems from the effect of boundaries on coherences. For larger $p$, these opposite momenta correlations eventually vanish very quickly in contrast to the weakly interacting case where they remain perfect due to condensate depletion by Bogoliubov quasi-particles pair creation. 

The probability distribution $p(n_p)$ of the number of bosons in a momentum state $p$ is shown to have a quasi-Gaussian shape at low momentum ($p$ smaller or of the order of the inverse radius of the gas $1/R$) and is exponential at larger momentum. Finite temperature causes the probability distribution for $n_p$ to evolve towards an exponential distribution. 

Finally, we discuss the rather academic problem of the distribution of the number of bosons in the fundamental natural orbital. The probability distribution of the quasi-condensate occupation was obtained numerically and resembles a truncated Gaussian curve. The probability of having zero bosons in the condensate is reduced compared to what happens without the harmonic trap. The presence of a trap is thus conducive to the formation of a quasi-condensate. However, in contrast to the two-dimensional case, no Bose-Einstein condensation is possible and the total number of bosons in the quasi-condensate still behaves as $\sqrt{N}$, where $N$ is the total number of bosons in the trap.

Our work is not only relevant for cold atom systems but also for some magnetic systems because the Tonks-Girardeau model is equivalent to the XXZ chain \cite{Caza2011,Guan2013}. One possible extension of this work is the study of the Lieb-Liniger model at strong but finite repulsion strength \cite{LiebLiniger,TanLiebLiniger,Felipe2019,Gangardt2003,GangardtShlyapnikov2,Nandanietal}. Non-equilibrium situation, such as the behavior after a quench could also be investigated. Coherences present in the density matrix, i.e. off-diagonal elements, may play a role \cite{Schehr2019}. They are usually neglected in the so-called generalized hydrodynamics approach to the time evolution of integrable systems \cite{JoelMoore}.


\section*{Acknowledgments} We would like to acknowledge helpful discussions with D. Cl\'ement, F. H\'ebert and A. Minguzzi.
 
\appendix

\section{Correlations in the homogeneous gas: effect of boundary conditions}
\label{app_hom_gas}

We briefly discuss the correlations between momentum occupation numbers for a homogeneous gas in a ring geometry (periodic boundary conditions) and in a box geometry (open boundary conditions). The procedure to compute $\mathcal G_{p,q}$ is the same as for a harmonic trap but the one-body density matrix are different. They are given by the following expressions \cite{Caza2011}

\begin{equation}
  \rho_1^R(x,y)=A_0 \rho_0 \left[\frac{\pi}{\rho_0 L \sin[\pi |x-y|/L]}\right]^{1/2},
  \label{eq_rho1_ring}
\end{equation}
and
\begin{equation}
  \rho_1^B(x,y)=B_0 \rho_0 \left[\frac{\rho_0^{-1}\sqrt{d(2x|2L)d(2y|2L)}}{d(x+y|2L)d(x-y|2L)}\right]^{1/2},
  \label{eq_rho1_box}
\end{equation}
with $A_0$ and $B_0$ some constants that are not needed for the computation of $\mathcal G_{p,q}$, $\rho_0$ the average density, $L$ the system size and $d(x|L)=L|\sin(\pi x/L)|/\pi$. In the thermodynamic limit ($L\to \infty$), $d(x|L)\to |x|$ and we retrieve the standard power-law correlations with exponent $1/2$.

\begin{figure} 
  \includegraphics[width=\linewidth]{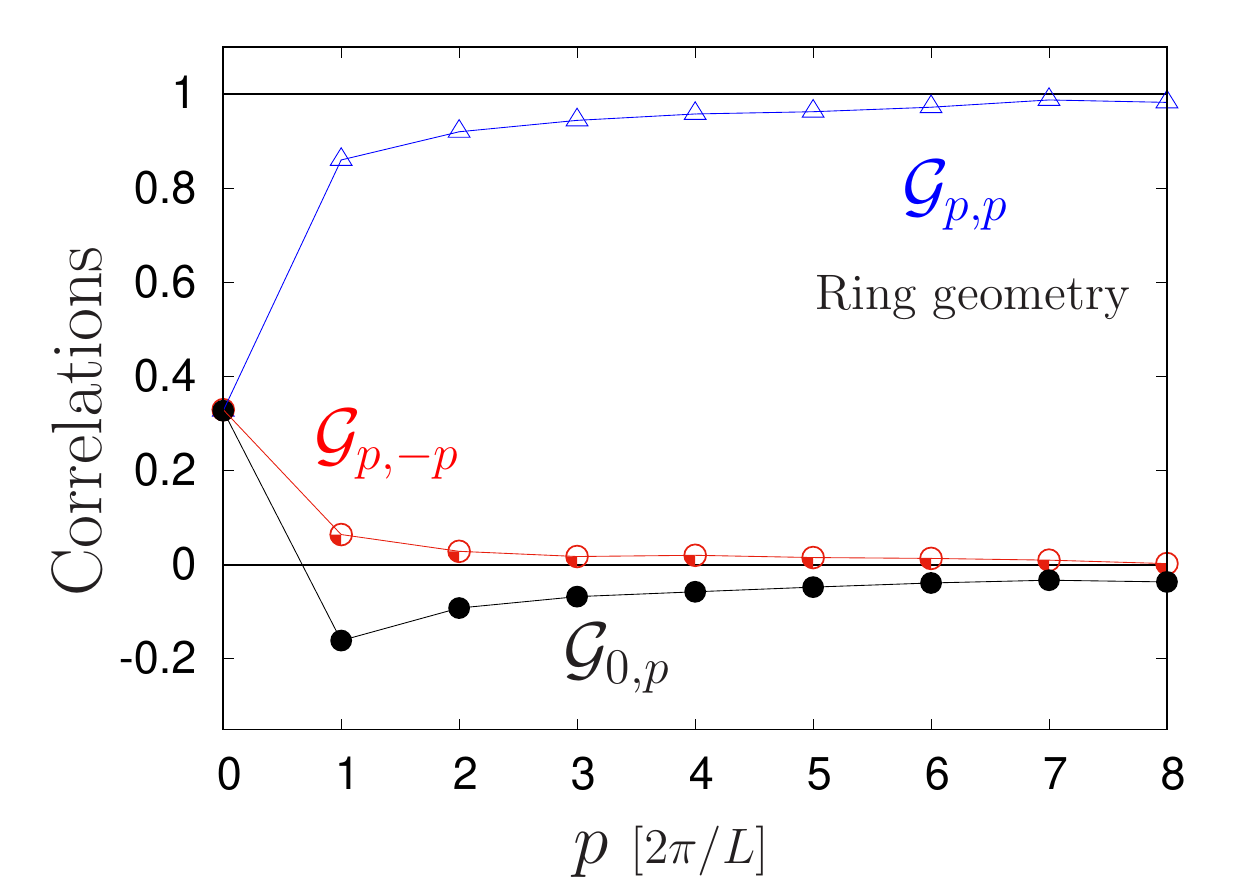}
  \caption{\label{fig_corr_ring} Cut of the momentum occupation number correlations $\mathcal G_{q,p}=\langle \hat n_p \hat n_q\rangle/\langle \hat n_p\rangle\langle \hat n_q\rangle-1$ along the lines $p=q$ (blue triangles), $q=0$ (black dots) and $p=-q$ (red circles) for an uniform Tonks-Girardeau gas in a ring geometry (periodic boundary conditions).}
\end{figure}

\begin{figure} 
  \includegraphics[width=\linewidth]{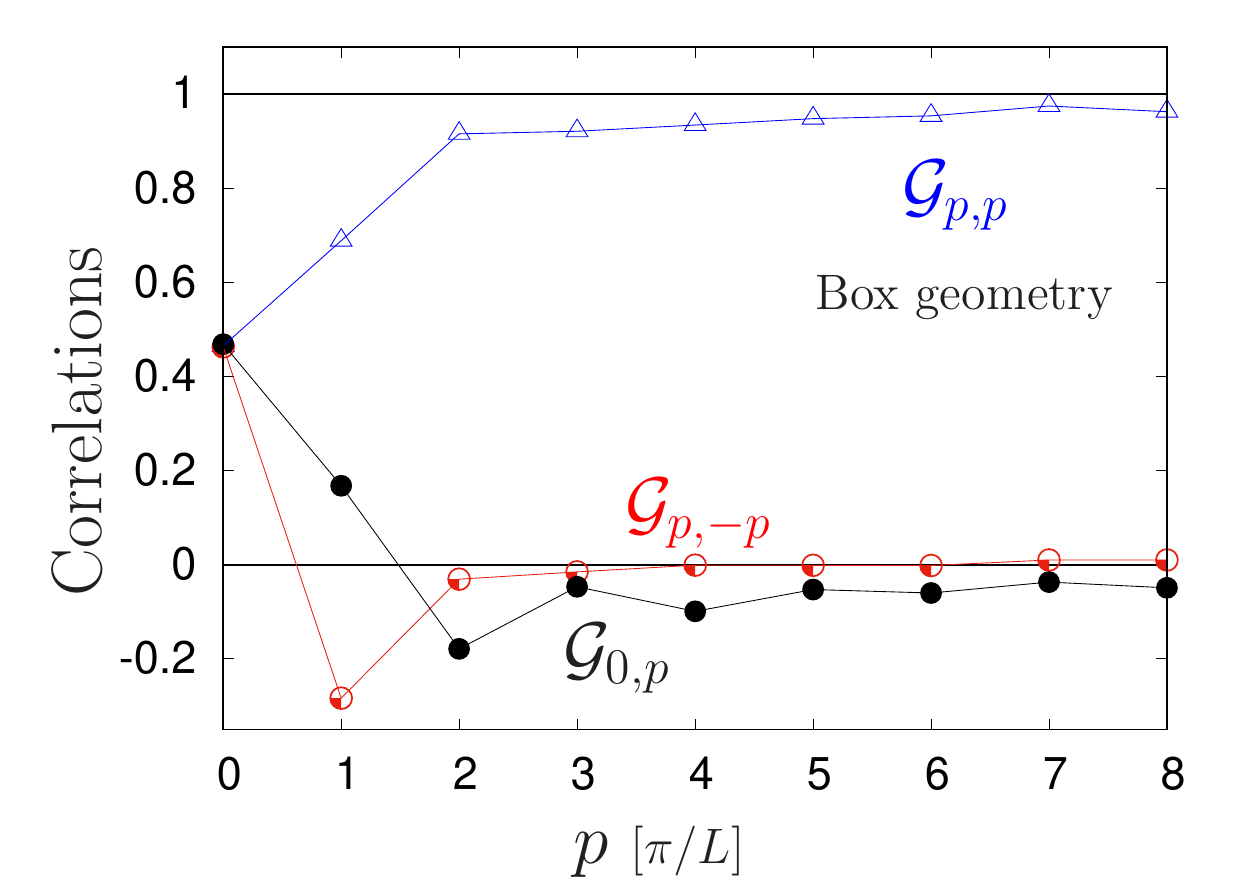}
  \caption{\label{fig_corr_box} Cut of the momentum occupation number correlations $\mathcal G_{q,p}=\langle \hat n_p \hat n_q\rangle/\langle \hat n_p\rangle\langle \hat n_q\rangle-1$ along the lines $p=q$ (blue triangles), $q=0$ (black dots) and $p=-q$ (red circles) for a Tonks-Girardeau gas in a box geometry (open boundary conditions).}
\end{figure}

As discussed in the main text, Figs. \ref{fig_corr_ring} and \ref{fig_corr_box} present similar features as in the case of a Tonks-Girardeau gas in a harmonic trap. The main difference is the negativity of $\mathcal G_{p,-p}$ for $pL$ ($pR$ for the trap) of order unity. This effect is totally absent in a ring geometry and therefore is related to boundary effects. To confirm this, we use the following analytical argument. $\mathcal G_{p,-p}$ is related to the Fourier transform of the two-particles density matrix
\begin{equation}
  \int dx_1 dx'_1 dx_2 dx'_2 \rho_2(x_1,x_2;x'_1,x'_2)e^{-i p(x_1-x'_1)}e^{i p(x_2-x'_2)}
\end{equation}
and is dominated by two contributions. The direct term, which corresponds to $x_1\simeq x'_1$ and $x_2\simeq x'_2$ which is positive and the  exchange term $x_1\simeq x'_2$ and $x_2\simeq x'_1$. The latter can be approximately cast into a term which is proportional to the integral
 \begin{eqnarray}
   &\,&  \int_0^L {e^{i p (x_1 - x^{\prime}_1)} \over \vert x_1 - x^{\prime}_1\vert^{1/2}} \, dx^{\prime}_1
   \int_0^L {e^{i p (x_2 - x^{\prime}_2)} \over \vert x_2 - x^{\prime}_2\vert^{1/2}} \, dx^{\prime}_2 \nonumber \\
   &\,&  \int_0^L\!\!\!\int_0^L \! n\Bigl(x_{CM} + {x_r \over 2}\Bigr) \,
   n\Bigl(x_{CM} - {x_r \over 2}\Bigr) \, e^{-2 i p x_r}  dx_r \, dx_{CM}, \nonumber \\
 \end{eqnarray}
 where $x_1$ ($x_2$) and $x^{\prime}_1$ ($x^{\prime}_2$) are the coordinates of the charges of the first (second) dipole. $x_{CM}$ is the center of mass of the two dipoles and $x_r$ their relative distance. This term can be negative and overcome the direct term. In a ring geometry, the density is constant and this integral is always null. This explains why $\mathcal G_{p,-p}$ is always positive in this case. In a box potential, the overlap of the densities is maximal when $x_r=L/2$ therefore the integral will be dominated by the neighborhood of this point. The integrand $\exp[-2 i p x_r]$ is maximally negative for $2 p x_r=\pi$, namely for $p=\pi/L$. This is exactly the value observed in Fig. \ref{fig_corr_box}. In the harmonic trap, the density profile $n(x)$ extending from $-R$ to $R$, is, to a good approximation proportional to $\vert 1 - x^2/R^2\vert^{1/2}$, see Eq. (\ref{eq_density_trap}). Taking a crude estimate of the profile by the characteristic function of the interval $\lbrack -R \, , \, R \rbrack$, the situation is similar to a box geometry of size $L=2R$. This predicts a mimimum $\mathcal G_{p,-p}$ around $pR=\pi/2\simeq 1.57$ which is consistent with Fig. \ref{fig_corr}. Similar arguments can be formulated for the oscillations in $\mathcal G_{0,p}$ (where there is no exchange term). Using the same procedure, it can be shown that the integrand is minimal for $p=2k\pi/L$ and maximal $p=(2k+1)\pi/L$ with $k$ an integer. In a ring geometry, the exchange term is thefore always negative whereas it oscillates in a box geometry.

\section{Finite temperature Lenard's expansion}
\label{app_lenard}

In this appendix, we derive the finite temperature Lenard expansion for the $n$-body density matrix for the trap. The only formal difference with the zero temperature case is that now, the kernel $K(x,y)$ has to be replaced by its finite temperature counterpart $K_T(x,y) \, = \, \sum_{\lbrace \nu_l\rbrace} \sum_{l=1}^N f_{\nu_l} u_{\nu_l}(x) u_{\nu_l}(y)$, with $f_{\nu}$ the Fermi function at energy $\epsilon_{\nu}$, $f_{\nu} = {1 \over e^{\beta \epsilon_{\nu}}+1}$ and $\beta = 1/(k_B T)$. The first sum runs over all possible sets $\lbrace \nu_1, ... ,\nu_N\rbrace$, all $\nu_l$, $l=1$ to $N$ being different from each other. For simplicity, we consider the direct term in $\langle n_p^2 \rangle$. The term with $n=2$ in Eq. (\ref{eq_lenard}) involves determinants like
\begin{eqnarray}
\left\vert
\begin{matrix}
K_T(x,y) & K_T(u,y) & K_T(x_1,y) & K_T(x_2,y) \cr
K_T(x,w) & K_T(u,w) & K_T(x_1,w) & K_T(x_2,w) \cr
K_T(x,x_1) & K_T(u,x_1) & \sqrt{2N} & K_T(x_2,x_1) \cr
K_T(x,x_2) & K_T(u,x_2) & K_T(x_1,x_2) & \sqrt{2N} 
\end{matrix}
\right\vert, \nonumber \\
\end{eqnarray}
since for all reasonable temperatures, $K_T(x,x) = \sqrt{2N}$. We assume that we take the direct term, $y$ close to $x$ and $w$ close to $u$; the two dipoles $(x,y)$ and $(u,w)$ being far apart. The intervals are $J_1 = \lbrack x \, , \, y \rbrack$ and $J_2 = \lbrack u \, , \, w \rbrack$. There are four ways of inserting $x_1$ and $x_2$ in the intervals $J_1$ and $J_2$. Let us suppose first that they are inserted in the same interval, $J_1$ for example. Then, $K_T(x_1, w) < {1 \over \pi \vert x_1 - w \vert}$ and $\vert x_1 -w\vert$ is in general of order $\sqrt{2N}/n$, with $n=2$ here. In contrast, $K_T(x_1,x)$ is of order $\sqrt{N}$. More generally, if $z_1$ and $z_2$ do not belong to the same dipole while $z^{\prime}_1$ and $z^{\prime}_2$ belong to the same dipole, the ratio $(K_T(z_1,z_2)/K_T(z^{\prime}_1,z^{\prime}_2))$ will be of order $1/N$ and can be neglected in the thermodynamic limit. Thus, if $x_1$ and $x_2$ belong to the dipole $(x,y)$, the determinant can be brought into the form
\begin{eqnarray}
\left\vert
\begin{matrix}
K_T(x,y)   & K_T(x_1,y)  & K_T(x_2,y)   & 0    \cr
K_T(x,x_1) & \sqrt{2N}   & K_T(x_2,x_1) & 0    \cr
K_T(x,x_2) & K_T(x_1,x_2) & \sqrt{2N}   & 0    \cr
0          & 0           & 0           & K_T(u,w) 
\end{matrix}
\right\vert, \nonumber \\
\end{eqnarray}
which factorizes into $K_T(u,w)$ times a function of the other variables. Taking the Fourier transform with respect to $(w-u)$ will not give any power law contribution in Tan contact's regime because $K_T(u,w)$ is an analytic function of $u$ and $w$. Thus, the points $x_1$ and $x_2$ must lie in two separate intervals, i.e. $x_1$ in $J_1$ and $x_2$ in $J_2$ or the reverse. We now examine these terms. Taking again into account the fact that the ratio $(K_T(z_1,z_2)/K_T(z^{\prime}_1, z^{\prime}_2))$ will be of order $1/N$ when $z_1$ and $z_2$ belong to different dipoles while $z^{\prime}_1$ and $z^{\prime}_2$ lie in the same dipole, in the thermodynamic limit, there is again a factorization of the 4 x 4 determinants and the corresponding term in the perturbation series can be put into the form $\int_{J_1} D_1(x,y;x_1) \, dx_1 \,\, \int_{J_2} D_1(u,w;x_2) \, dx_2 $ with
\begin{eqnarray}
D_1(x,y;x_1) \, = \, 
\left\vert
\begin{matrix}
K_T(x,y) & K_T(x_1,y) \cr
K_T(x,x_1) & \sqrt{2N},
\end{matrix}
\right\vert.
\end{eqnarray} 
The contribution which gives the dominant term in Tan contact's regime will take the form ${(-2)^2 \over 2!}$ $sgn(u-x)$ $sgn(w-y)$ $\int_{J_1} D_1(x,y;x_1) \, dx_1 $ $\int_{J_2} D_1(u,v;x_2) \, dx_2$. Thus, there is a factorization into two distinct contributions. In the thermodynamic limit and in the regime of the contact, to lowest order in $1/p$, the equality $\langle n_p^2 \rangle \, = \, 2 \, \langle n_p\rangle^2$ holds.

\section{Correction to $ \mathcal G_{p,q}$ in the contact regime}
\label{app_tan}

We provide an estimate of the leading order correction to $ \mathcal G_{p,q}$ in the contact regime. If the dipoles $(x,y)$ and $(u,w)$ are not close to each other, there are nonzero terms which are of order $1/N$ at least; their contribution to $\mathcal G_{p,q}$ is zero in the thermodynamic limit.  The main term will come from the configurations where the dipoles $(x,y)$ and $(u,w)$ are close to each other. Let us look at the direct term; $x$ and $y$ are typically $1/p$ apart whereas $u$ and $w$ are $1/q$ apart. We take $x=0$ for simplicity and $q$ very close but not equal to $p$. On integrating over $u$, the only region which will make a contribution will be for $u$ of the order of $(1/p + 1/q)$ otherwise dipoles do not interact. Now, the points $x,y,u,w$ are all within a range of $1/p$ and we can no longer make the simplifications mentioned in Appendix \ref{app_lenard}. To get a nonzero contribution, we need a non-analyticity \cite{Lighthill} in the behavior of $\rho_B(x,u;y,w)$ as a function of the variables $x$, $u$, $y$, and $w$. In Lenard's expansion, Eq. (\ref{eq_lenard}), the term with $n=1$ involves the introduction of only one extra point $x_1$ and the computation of a  three by three determinant. However, this turns out not to be sufficient because we need non-analyticity in both $y$ and $(w-u)$. Therefore, the first term in the expansion of $\rho_B(x,u;y,w)$ which will give a nonzero contribution to $\mathcal G_{p,q}$ will be the term $n=2$ and two extra points $x_1$ and $x_2$ have to be introduced. The resulting four by four determinants are zero as soon as, among the variables $y$, $w$, $x_1$ or $x_2$, two of them are equal because two lines of the determinant would be identical. Likewise, the determinant is also zero if among the variables $u$, $x$, $x_1$ or $x_2$, two of them are equal because two columns of the determinant would be the same. Therefore the determinants carry a factor $B$, with $B =  (y-x)(x-x_1)(y-x_2)(w-x_1)(w-x_2)(x_1-x_2)u(u-x_1)(u-x_2)x_1(x_1-x_2)x_2$. Notice that the dependence is in $(x_1-x_2)^2$. Upon integration over $x_1$ and $x_2$ in two intervals $J_1$ and $J_2$, and subsequent integration over $y$ and $(w-u)$, the overall factor goes as $p^{-16}$. Integration over $u$ brings an extra factor $2/p$ and dividing by $\langle n_p\rangle \langle n_q \rangle$ to obtain $\mathcal G_{p,q}$ gives a factor $p^8$. Overall, $\mathcal G_{p,q}$ is proportional to $p^{-9}$, (provided $p \not = q$, otherwise $\mathcal G_{p,p}=1$ due to the exchange term). A generalization of the argument to $q$ very different from $p$ leads to $\mathcal G_{p,q}$ being proportional to $(p^{-1} + q^{-1}) p^{-4}q^{-4}$, as stated in the main text.

\section{Probability density reconstruction from a finite number of moments}
\label{app_proba_reconstruct}

Reconstructing the probability density from its moments is an important and non-trivial mathematical problem \cite{Stieltjesmomentproblem,BarrySimon,theGreeks}. We present here a simple method which is well suited to our problem that do not suffer form mathematical pathologies. Moreover, we have checked that this method is fully consistent with other methods such as the maximum entropy method \cite{maximumentropyreconstruction} or the method of orthogonal polynomials \cite{orthogonalpoynomialsreconstruction}.

The procedure is rather simple. We take an ansatz for the probability distribution of the form

\begin{equation}
  p(x,\{a_i\})=\sum_{i=0}^{M-1} a_i x^i \,g(x),
\end{equation}
with $M$ the number of known moments and $g(x)$ a suitable decaying function. In our case, $g(x)$ has been taken to be $\exp(-x)$ or $\exp(-x^2)$ with similar accuracy. Then, we formally compute the moments of this trial distribution

\begin{equation}
  \langle x^k\rangle=\int_0^\infty x^k p(x,\{a_i\}).
\end{equation}
The $M$ coefficients of the polynomial are then the solutions of the linear system of $M$ equations $\langle x^k\rangle=\mathcal M_k$, with $\mathcal M_k$ the moments of the target distribution.

\section{One-particle density matrix at finite temperature}
\label{app_rho1_T}

We explain in this appendix how to compute the correlations at finite temperature in the hydrodynamic regime. We use a local density approximation (LDA) and bosonization on the one-particle density matrix and compare our formula to an exact calculation with $N=10$ bosons in the Tonks-Girardeau limit. The higher-order density matrices are then computed with the Wick's theorem (\ref{eq_wick}).

\begin{figure} 
  \includegraphics[width=\linewidth]{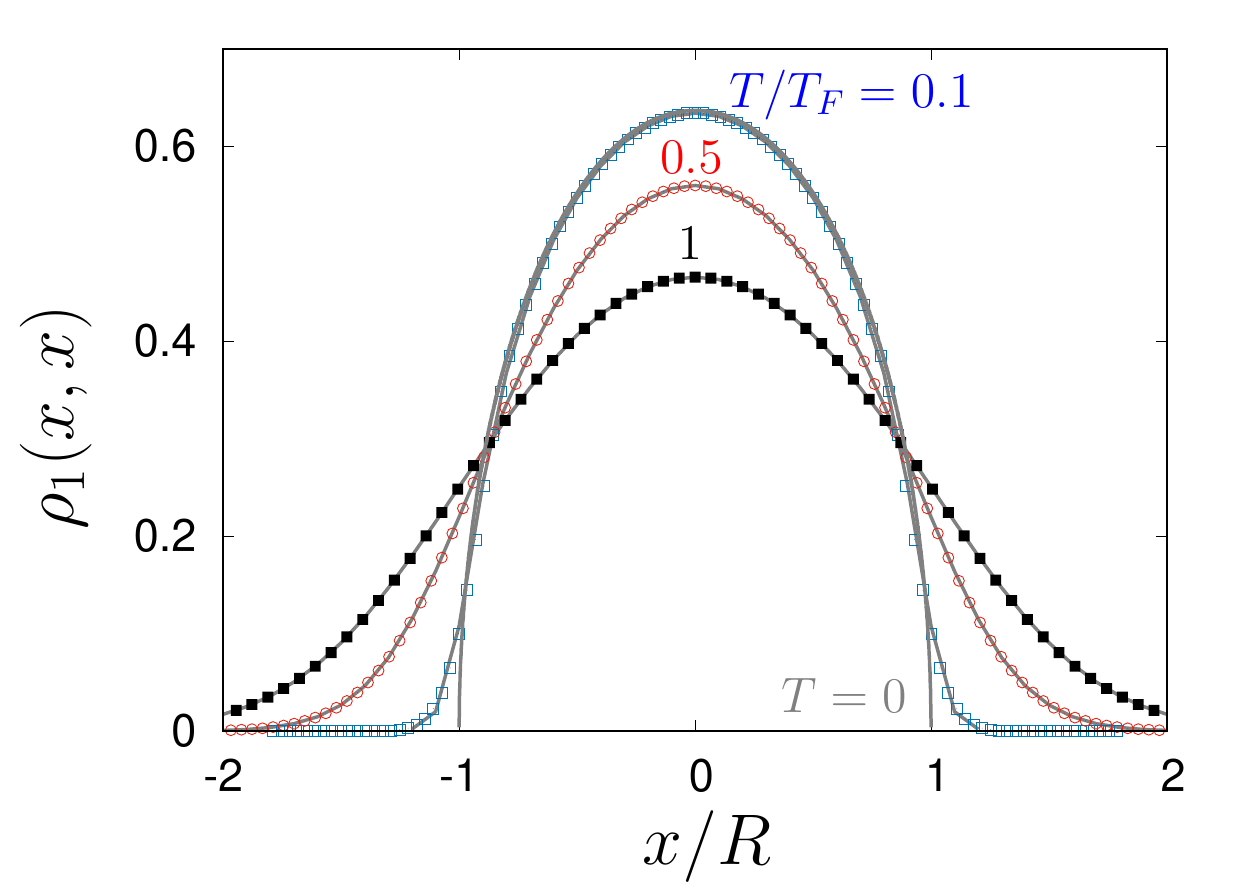}
  \caption{\label{fig_rho1} Diagonal part of the one-particle density matrix $\rho_1(x,x)$ in real space for different temperatures (density profile). Symbols are exact calculations with $N=10$ bosons and thin gray lines the corresponding prediction of formula (\ref{eq_nT}). The Fermi temperature is defined as $k_B T_F=E_F=N\hbar\omega$. The radius of the cloud at $T=0$ is $R=\sqrt{2N}a_0$.}
\end{figure}
\begin{figure} 
  \includegraphics[width=\linewidth]{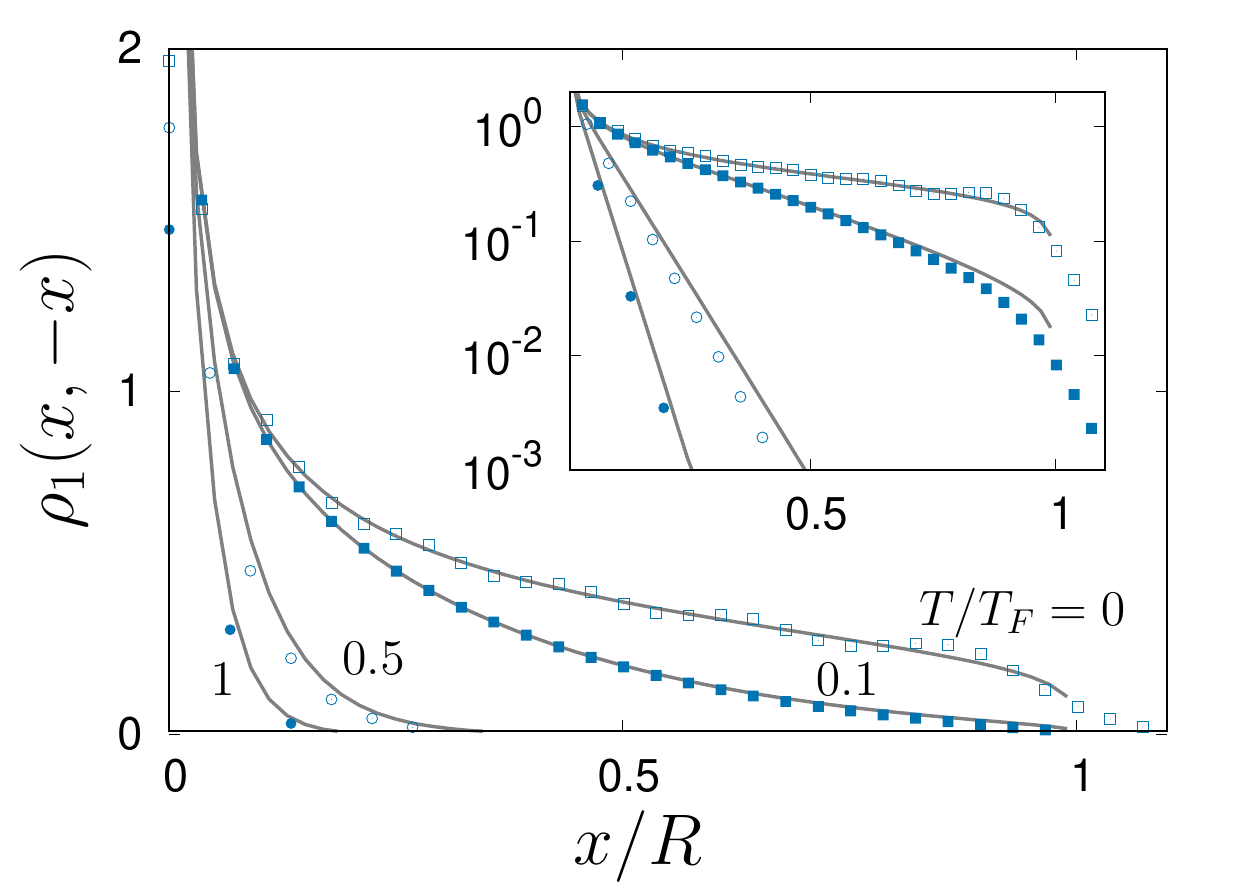}
  \caption{\label{fig_rho1_of} Off diagonal part of the one-particle density matrix $\rho_1(x,-x)$ in real space for different temperatures. Symbols are exact calculations with $N=10$ bosons and thin blue lines the corresponding prediction of formula (\ref{eq_rho1_T}). The Fermi temperature is defined as $k_B T_F=E_F=N\hbar\omega$. The inset shows the same data in logarithmic scale.}
\end{figure}

Finite temperature will induce two important corrections to the one-body density matrix. First, the density profile will be modified and evolves from the semi-circle law at $T=0$ to a Gaussian profile at high temperature. This modification of the density profile is exactly the same as for a gas of free fermions and has been shown to be accurately described by the following formula \cite{Dean2016}

\begin{eqnarray}
  n_T(x) \, &=& - {1 \over a_0 \sqrt{N}} \sqrt{{k_B T \over 2 \pi N \hbar \omega}} \nonumber \\
  &\textrm{Li}_{1/2}&\biggl(-(e^{ {N \hbar \omega \over k_B T}}-1)\exp\left[-\frac{m\omega^2 x^2}{2k_B T}\right]\biggr),
  \label{eq_nT}
\end{eqnarray}
where $\textrm{Li}_{1/2}$ is the polylogarithm function \cite{Grasd}. However, this correction is not important at low temperature ($kT\le 0.1 E_F$, $E_F=N\hbar\omega$) as shown on Fig. \ref{fig_rho1}. The most important correction is in the off-diagonal part of the density matrix, where real space correlations decay algebraically at zero temperature but exponentially at finite temperature. This decay is governed by the thermal length $\ell_T$ which is related to the Fermi velocity. In the homogeneous case, the one-body density matrix is obtained by substituting the term $1/\sqrt{|x-y|}$ by $[\ell_T \sinh(\pi |x-y|/\ell_T)]^{-1/2}$, with $\ell_T=\hbar v_F/k_B T$ \cite{Caza2004}. However, in the presence of a trap, the Fermi velocity becomes inhomogeneous. As a first approximation, we take its value at the center of the trap. Then, the one-body density matrix takes the form

\begin{equation}
  \rho^H_{1,T}=\frac{\mathcal{A}\, [n_T(x) n_T(y)]^{1/4}} {[\ell_T \sinh(\pi |x-y|/\ell_T)]^{1/2}},
  \label{eq_rho1_T}
\end{equation}
with $\mathcal A$ a constant that is not needed for the calculation of $\mathcal G_{p,q}$. In a harmonic trap, the Fermi velocity at the center is given by $v_F=\hbar\pi n_T(0)/m$. It therefore scales with $\sqrt{N}$ and not $N$ as the Fermi energy does. Finite temperature corrections will be crucial when the thermal length becomes smaller than the cloud radius $R= \sqrt{2N}a_0$. The ratio between these two quantities is easily calculated and is given by Eq. \ref{eq_LT} of the main text. As $N$ is increased, it is then possible to completely destroy phase coherence in the cloud while keeping a zero temperature density profile. The accuracy of this simple formula is checked by comparing this prediction to an exact calculation, based on the method developed in \cite{Atas2017}, with $N=10$ bosons as can be seen on Fig. \ref{fig_rho1_of}.

\end{document}